# Is the Difference Between the Pion Form Factor Measured in $e^+e^-$ Annihilations and $\tau^-$ Decays Due to an $H^-$ Propagator?

William M. Morse

*Physics Department, Brookhaven National Lab, Upton, NY 11973*
*(E-mail: Morse@bnl.gov)*

October 4, 2004

We discuss how a charged Higgs propagator would modify the form factor in $\tau^- \rightarrow \pi^- \pi^0 \nu_\tau$ decays.

Ref.1 compared the pion form factors available from $e^+e^- \rightarrow \pi^+ \pi^-$ (CMD2 experiment) and $\tau^- \rightarrow \pi^- \pi^0 \nu_\tau$ (ALEPH, CLEO, and OPAL experiments) in the range $0.37 < s < 0.92\ GeV^2$ where $s = m^2_{\pi\pi}$. The evaluation used CVC with estimates of isospin violation between the charged and neutral states for the isovector part. This mass range is dominated by the $\rho(770)$ resonance. Generally agreement was found within several percent, except for the higher masses, where the discrepancy was about 10%. It was pointed out that some of the discrepancy may be due to different masses and widths of the charged and neutral $\rho$ mesons [2,3]. Recently the KLOE experiment [4] has measured the pion form factor through the radiative process $e^+e^- \rightarrow \pi^+ \pi^- \gamma$ in the range $0.35 < s < 0.94\ GeV^2$. The situation has been reviewed [5] by Andreas Hocker at ICHEP 04, Beijing, Aug 16-22, 2004. He concludes that even after correcting for possible differences in the masses and widths of the charged and neutral $\rho$ mesons, there is still a discrepancy at the higher masses of 5-10%. Fig. 1 shows the pion form factor measured by the KLOE experiment [4]. The statistical errors range from ±2% at low mass to ±0.6% at higher masses. The systematic error is ±1.3%.

The decay $\tau^- \rightarrow \pi^- \pi^0 \nu_\tau$ can proceed through either $W^-$ exchange or $H^-$ exchange [6]:

$$\psi^2_{\pi\pi\nu} = (\psi_W + \psi_H)^2 = \psi_W^2 + 2\psi_W\psi_H + \psi_H^2 \approx \psi_W^2 + 2\psi_W\psi_H \quad (1)$$

The $W^-$ diagram is dominated by the $\rho(770)$ resonance in the mass range $0.35 < s < 0.94\ GeV^2$. There is no $\pi\pi$ resonance with the quantum numbers of the $H^-$ in this mass range. The hadronic vector current [7] is given by:

$$J^V_h = \sqrt{2} \cos\theta_c\, F_\pi(s)(q_{\pi^-} - q_{\pi^0}) \quad (2)$$

Fig. 1. Some representative points from the pion form factor $|F_\pi(s)|^2$ vs. $s$ (GeV$^2$) measured by the KLOE experiment [4]. The interference between the broad $\rho(770)$ and the narrow $\omega(782)$ can be seen. The statistical errors range from ±2% at low mass to ±0.6% at higher masses.

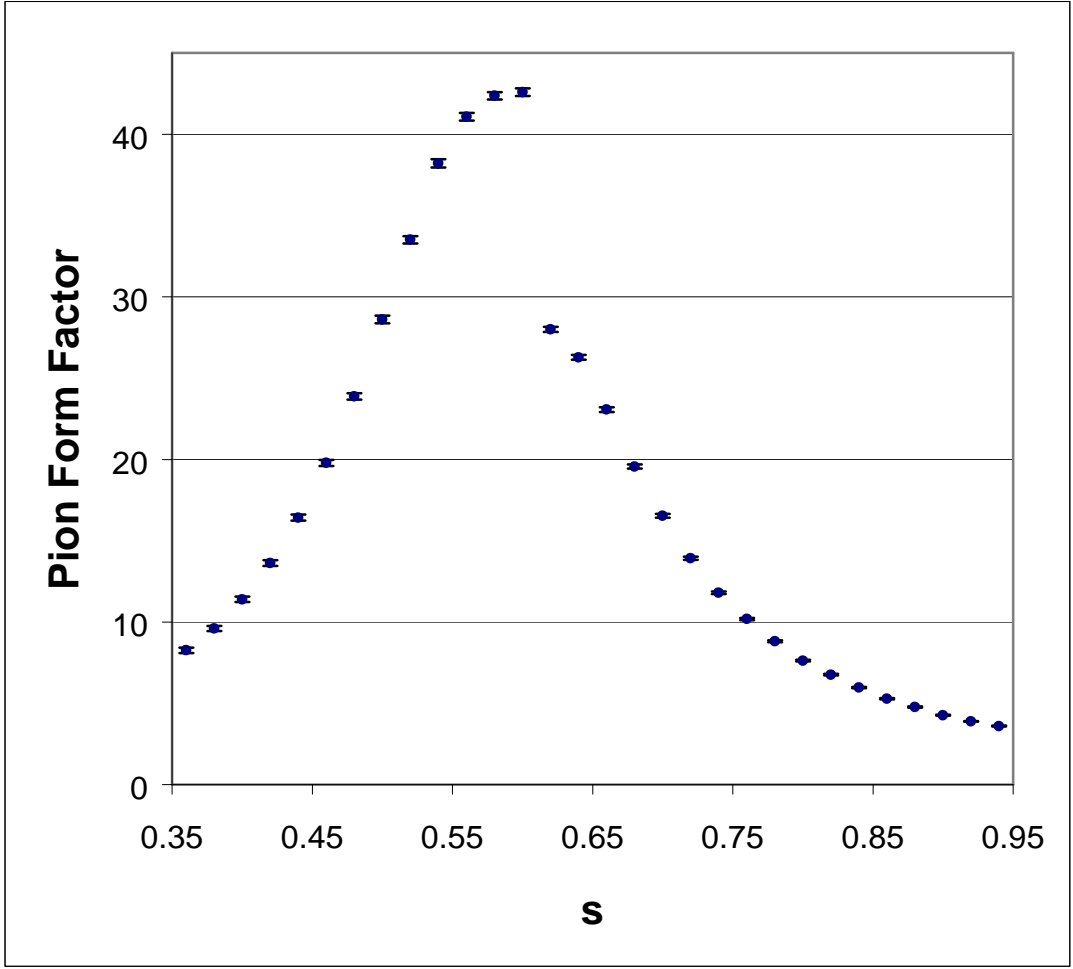

$F_\pi(s)$ is dominated by the $\rho^-(770)$ resonance in the mass range $0.35 < s < 0.94 \ GeV^2$. The two pions are in a $P$ wave. The hadronic scalar current is given by:

$$J_h^S = f_S(s) \qquad (3)$$

where the two pions are in a $S$ wave. We parameterize the $s$ dependence of $F_\pi$ and $f_S$ over this mass range by Breit-Wigner amplitudes:

$$\psi_W \propto \frac{m_\rho \Gamma_\rho}{m_\rho^2 - s - im_\rho \Gamma_\rho} \qquad \psi_H \propto \frac{A_H}{m_H^2 - s - im_H \Gamma_H} \approx \frac{A_H}{m_H^2} \qquad (4)$$

where for this study, we have ignored the $\rho(1450)$ [8] etc. We show in Fig. 2

$$R = \frac{|\psi^2_{\pi\pi\nu}| - |\psi^2_W|}{|\psi^2_W|} \quad (5)$$

as a function of *s*. This shape is in agreement with the difference of the pion form factor measured in $e^+e^-$ annihilations and $\tau$ decays within the uncertainties. This is our main result.

For the $\tau^-$ semi-leptonic decay model discussed in ref. 6, the additional $H^-$ couplings are given by:

$$g^S = -\frac{m_\tau(m_u + m_d)}{m^2_{H^-}}\tan^2\beta \quad (6)$$

With the minus sign in equ. 6, we get that the $\tau$ form factor should be higher than the $e^+e^-$ form factor above the $\rho$ mass, in agreement with ref. 1. Ref. 9 gives the limit

$$\frac{\tan\beta}{m_{H^-}} < 0.4\ GeV^{-1} \quad (7)$$

coming mainly from the *B* leptonic and semi-leptonic decays. The direct limit [9] on the mass of the charged Higgs is 79.3 *GeV*. If we use the "current quark mass" of $m_u + m_d = $ 6 - 12 *MeV* [9], we get $|g^S| < 0.0034$.

Fig. 2. *R* in arbitrary units as a function of *s* (GeV$^2$). This shape is in agreement with the difference of the pion form factor measured in $e^+e^-$ annihilations and $\tau$ decays within the uncertainties. This is our main result.

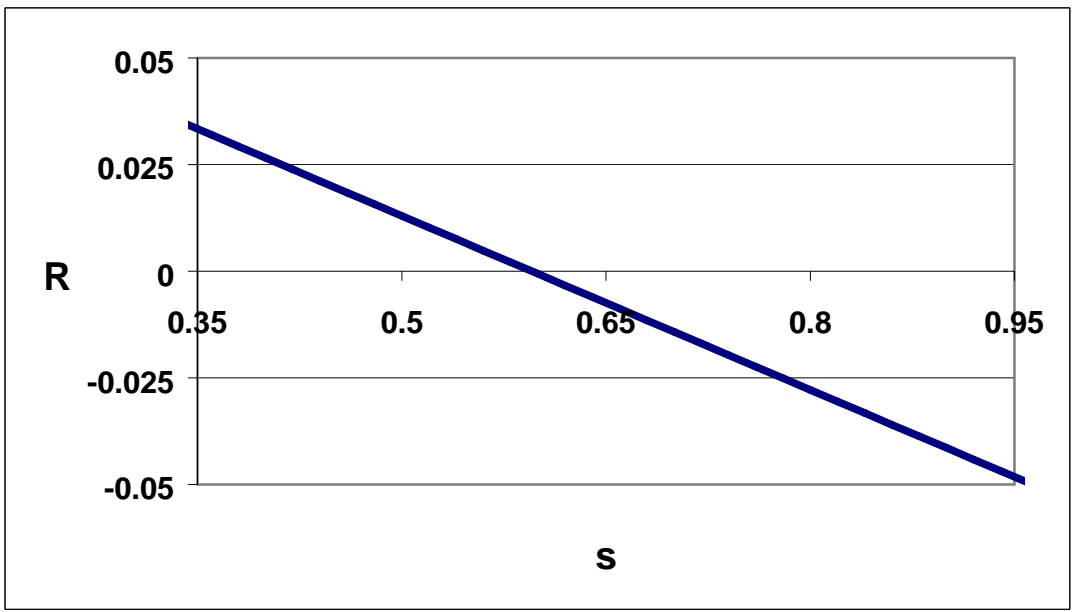

From above, we get:

$$R \cong 2g^S \frac{(m_\rho^2 - s)}{m_\rho \Gamma_\rho} \quad (8)$$

which gives at $s \approx 1 \ GeV^2$, $|R| \leq 2.5\%$.

In conclusion, we have shown that the disagreement between the measured $\tau^-$ and CVC predicted pion form factors is consistent with the interference of the $W^-$ and $H^-$ diagrams in shape. This is our main result. A quick estimate of the allowed size of the effect is given for a model discussed in ref. 6, which gives about one half the observed difference in the form factors.

This work was supported in part by the United States Department of Energy. I wish to acknowledge interesting conversations with W. Marciano.